\documentclass[12pt]{spieman}  		
\usepackage{amsmath,amsfonts,amssymb}
\usepackage{graphicx}
\usepackage{setspace}
\usepackage{tocloft}

\def\wisk#1{\ifmmode{#1}\else{$#1$}\fi}
\def\deg    {\wisk{^\circ}}
\def\amin   {\wisk{^\prime\ }}

\title{Polarized Beam Patterns from a Multi-Moded Feed
For Observations of the Cosmic Microwave Background}

\author[a,*]{A. Kogut}
\author[a,b]{D. J. Fixsen}
\affil[a]{Goddard Space Flight Center, Code 665, 
	8800 Greenbelt Road,
	Greenbelt, MD, USA 20771}
\affil[b]{University of Maryland, College Park, MD, USA 20742}

\cftpagenumbersoff{figure}
\cftpagenumbersoff{table} 
\begin{document} 
\maketitle

\begin{abstract}
We measure linearly polarized beam patterns
for a multi-moded concentrator
and compare the results to a simple model
based on geometric optics.
We convolve the measured co-polar and cross-polar beams
with simulated maps of CMB polarization
to estimate the amplitude of the systematic error
resulting from the cross-polar beam response.
The un-corrected error signal has typical amplitude of 3 nK,
corresponding to inflationary B-mode amplitude $r \sim 10^{-3}$.
Convolving the measured cross-polar beam pattern
with maps of the CMB E-mode polarization
provides a template for correcting the cross-polar response,
reducing it to negligible levels.

\end{abstract}

\keywords{beam pattern, polarimeter, cosmic microwave background}

{\noindent \footnotesize\textbf{*}
Address all correspondence to:
Alan Kogut,  \linkable{alan.j.kogut@nasa.gov} }

\begin{spacing}{2}   		

\section{Introduction}
\label{sect:intro}  		

Linear polarization of the cosmic microwave background
provides a unique window
into the physics of the early universe.
Gravitational waves
created during an inflationary epoch
interact with the CMB at later times
to impart a distinctive signature in linear polarization.
For the simplest (single-field) inflation models,
the amplitude of this B-mode signal 
depends on the energy scale of inflation as
\vspace{-1mm}
\begin{equation}
E = 1.06 \times 10^{16} 
\left( 
\frac{r}{0.01} 
\right)^{1/4}
~{\rm GeV} 
\label{potential_eq}
\end{equation}
where 
$r$ is the power ratio of gravitational waves to density fluctuations
\cite{lyth/riotto:1999}.
Current measurements set upper limits $r < 0.07$
at 95\% confidence
\cite{bicep2xkeck_2016}.
If inflation results from Grand Unified Theory physics
(energy $E ~ \sim ~10^{16}$ GeV),
the B-mode amplitude should be of order 30 nK.
Signals at this level
could be detected by  a dedicated polarimeter,
providing a critical test of a central component of modern cosmology.

The inflationary signal is faint compared to the
fundamental limit imposed by photon noise statistics
for a single-moded detector in a one-second integration.
Even ideal (noiseless) detectors suffer from this limit;
the only solution is to collect more photons.
The light-gathering ability of a detector can be specified by its etendu
$A \Omega$,
where $A$ is the detector area
and $\Omega$ is the solid angle.
Increasing the etendu adds modes,
thereby increasing both the signal and the photon  noise.
The modes are independent,
so that the noise increases as the square root of the etendu
${\rm NEP} \propto (A \Omega)^{1/2}$.
But since the signal increases linearly with etendu,
the signal-to-noise ratio improves as $(A \Omega)^{1/2}$,
increasing the overall system sensitivity.

A common implementation for CMB polarization
increases the instrument etendu
using a focal plane tiled 
with thousands of detectors
(for a recent review, see 
[\citenum{cmb_s4_technology_2017}]).
These systems typically
couple the detector to the sky
using optical structures
(feed horn, lenslet, phased antenna array)
which restrict the system response
to a single electromagnetic mode at the sensor.
Single-moded systems achieve
diffraction-limited angular resolution
with well-defined (Gaussian) beam profiles,
but the large number of sensors required
drives system-level complexity and cost,
both directly
and through the accompanying need
for cryogenic multiplexing to reduce the wire count
to cold stages of the instrument.

Multi-moded sensors provide an alternative design solution.
The beam width of a diffraction-limited system
observing a single polarization state in a single mode
scales with wavelength,
${A \Omega=\lambda^2}$.
In a multi-moded system,
the etendu is fixed so that the 
number of modes scales as
${N_{\rm mode} = {A \Omega} / {\lambda^2}}$.
The different electromagnetic modes
form an orthogonal basis set
and add incoherently at the detector.
Assuming nearly equal occupancy for most modes, 
the detector sensitivity then scales as $N_{\rm mode}^{1/2}$.
For detector area {$A ~\gg~\lambda^2$},
the number of modes is large,
allowing a corresponding
reduction in detector count
compared to a single-moded system of comparable sensitivity.

\begin{figure}[b]
\begin{center}
\includegraphics[height=3.5in]{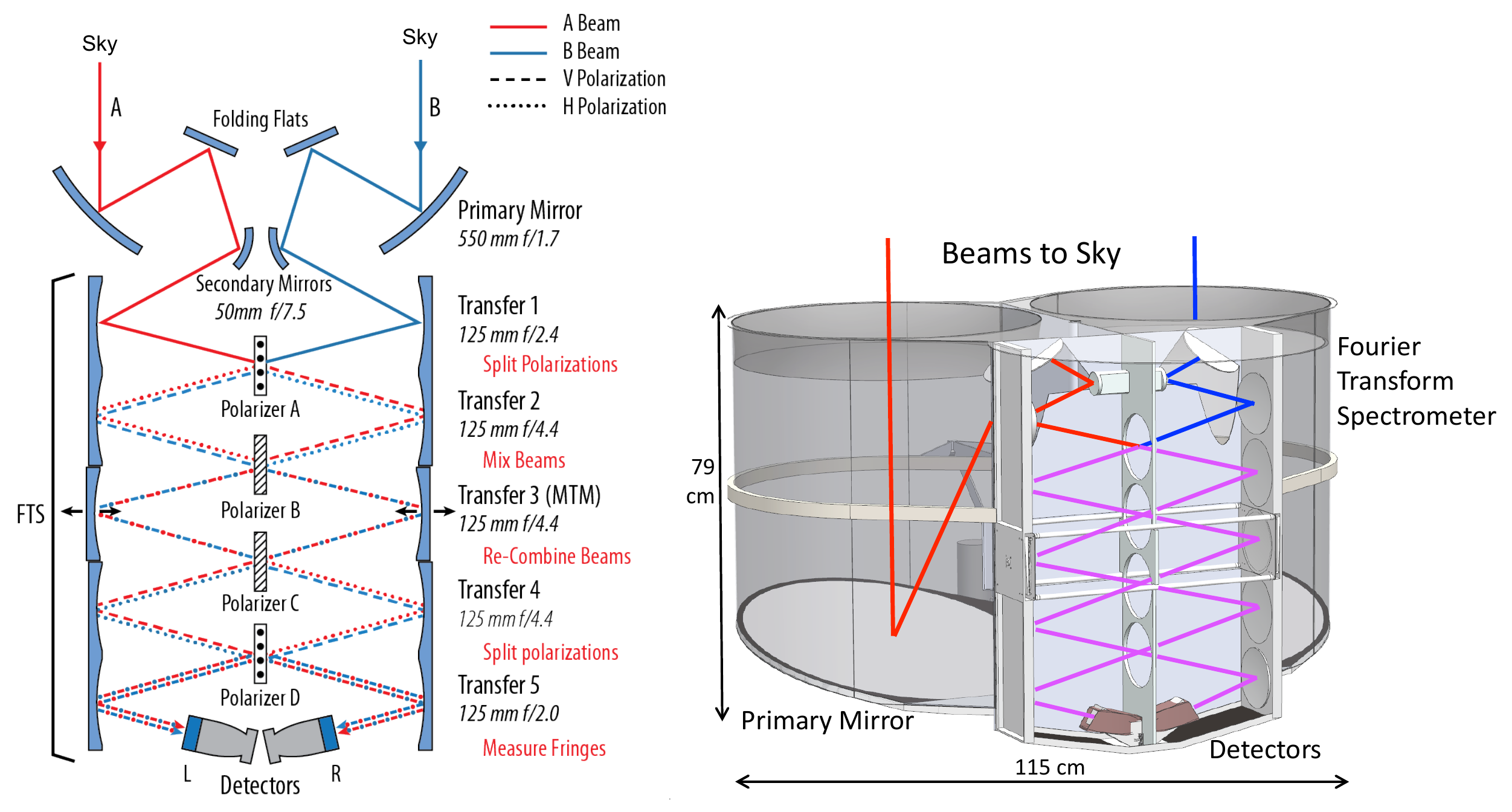}
\end{center}
\caption[PIXIE optical signal path]
{
(Left) Schematic view of the PIXIE optical signal path.
A polarizing Fourier transform spectrometer
interferes the signal from two co-pointed beams
to produce a fringe pattern proportional
to the difference spectrum
between orthogonal polarization states
from the two beams.
Polarization-sensitive detectors
mounted within multi-moded concentrators
sense the fringe pattern.
(Right) Physical layout of PIXIE optics.
}
\label{pixie_schematic}
\end{figure}

The Primordial Inflation Explorer (PIXIE)
\cite{kogut/etal:2011,
kogut/etal:2016}
is a mission concept to measure CMB polarization
at levels $r < 10^{-3}$.
PIXIE combines four multi-moded detectors
with a polarizing Fourier transform spectrometer (FTS)
to achieve background-limited sensitivity
across a broad frequency range.
Figure \ref{pixie_schematic} shows the optical design.
A pair of primary mirrors 55 cm in diameter
produce two co-aligned beams on the sky.
Folding flats and secondary mirrors
transfer the beams to the FTS,
which mixes the beams
and introduces an optical phase delay.
The recombined beams are then routed
to a pair of non-imaging concentrators,
each of which contains
two multi-moded bolometric detectors
to sense the fringe pattern 
as a function of optical phase delay.
The instrument etendu of 400 mm$^2$ sr
corresponds to flat-topped beam with 
full width at half maximum 2.6\deg~on the sky,
and is conserved throughout the optical path.

An important aspect of the proposed instrument design
is the use of non-imaging concentrators
to couple light from the FTS to the detectors.
Each detector is sensitive to a single linear polarization;
any cross-polar response to the orthogonal polarization
produces an effective rotation of the measured polarization
with respect to the true polarization on the sky.
The cross-polar response
of single-moded structures is well understood.
A corrugated circular feedhorn,
for example,
will typically exhibit
cross-polarization at levels -25 dB
with a distinctive quadrupolar (cloverleaf) angular pattern
\cite{kildal:2000}.
The cross-polar response of multi-moded structures
is more complicated.
We describe measurements and modeling
of the co-polar and cross-polar beam patterns
for a rectangular multi-moded concentrator
designed for the PIXIE polarimeter,
and assess the impact of the measured cross-polar response
for measurements of the inflationary polarization signal.

\begin{figure}[b]
\centerline{
\includegraphics[height=7cm]{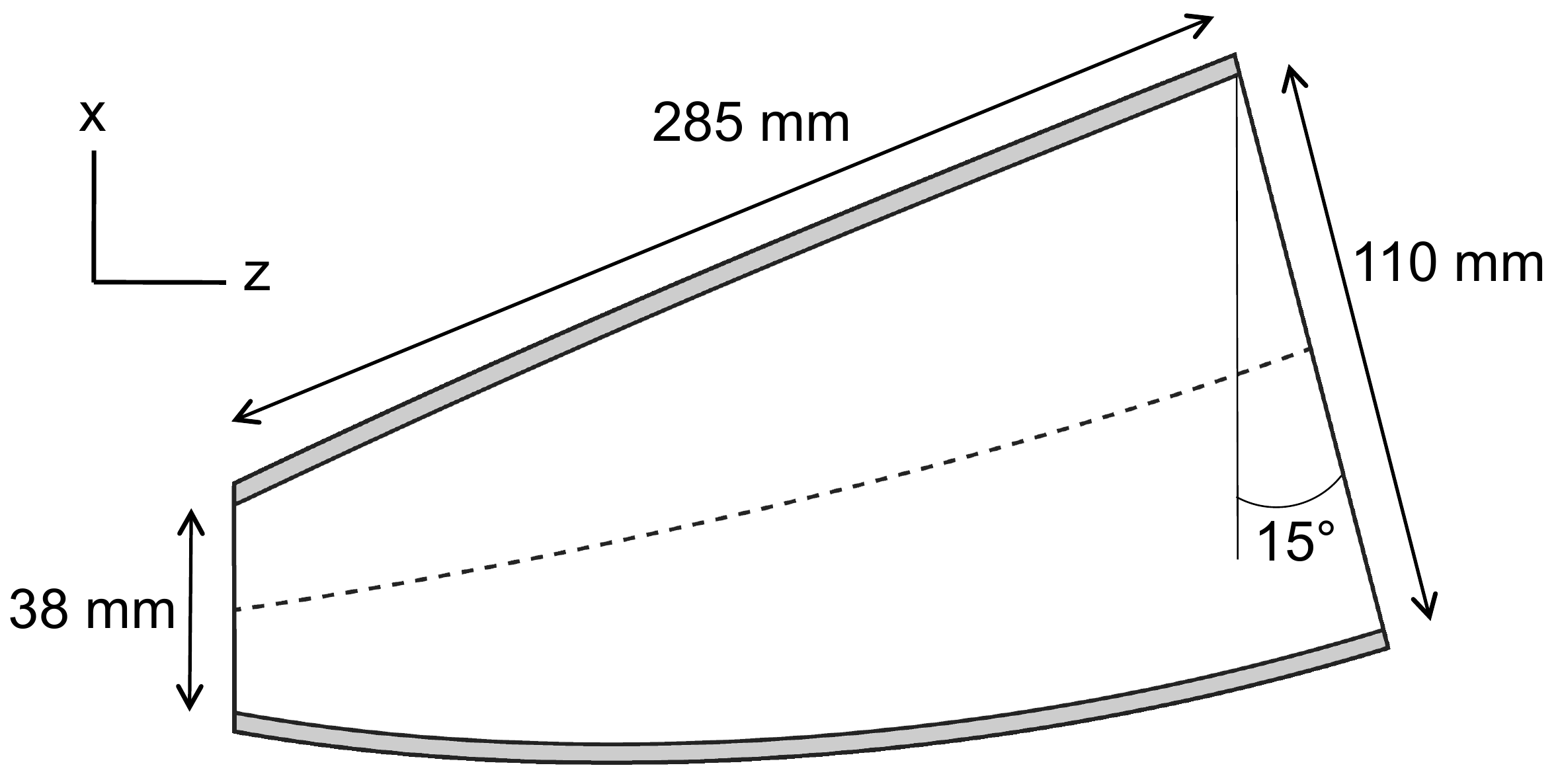}
}
\caption[Schematic of PIXIE concentrator]
{
Schematic of the rectangular concentrator 
in the $\hat{x}-\hat{z}$ plane.
The dashed line indicates the junction of the 
$\hat{u}$ and $\hat{v}$ walls,
which project out of the plane of the page 
(see text).
Dimensions are for the larger scaled concentrator
used to measure the beam patterns;
the concentrator for the actual instrument 
is 3 times smaller in all dimensions.
}
\label{feed_schematic}
\end{figure}

\begin{figure}[b]
\centerline{
\includegraphics[height=7cm]{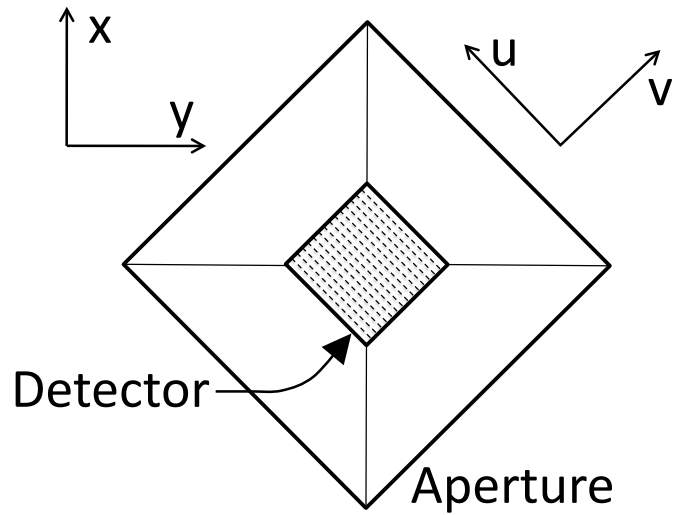}
}
\caption[Instrument and concentrator coordinate systems]
{
Schematic view looking into the concentrator aperture
from the FTS,
showing the instrument ({\it xy})
and concentrator ({\it uv})
coordinate systems.
The concentrator is rotated 45\deg~
with respect to instrument coordinates
so that asymmetries in instrument coordinates
affect both detectors equally.
Dashed lines show the orientation of the absorbing grid
for the $\hat{u}$ detector.
}
\label{uv_schematic}
\end{figure}

\section{Multi-Moded Concentrator Design}
\label{sect:concentrator}  
Figure \ref{feed_schematic} shows the concentrator design,
which transforms the $f/2$ beam from the FTS
to a $2\pi$~sr solid angle at the detector.
To preserve linear polarization,
the concentrator has a rectangular cross section,
with 4 walls
each of which forms an elliptical section.
Each concentrator contains
a pair of polarization-sensitive bolometers,
mounted within an integrating cavity
at the exit aperture of the concentrator.
Each bolometer detects a single linear polarization;
the pair are mounted with their polarization directions
rotated by 90\deg
~to measure orthogonal polarization states.

Light from the FTS enters the concentrator
at an angle of 15\deg ~from the $\hat{z}$ axis.
The concentrator design minimizes 
the resulting asymmetry in the polarized beam patterns
by rotating the concentrator 45\deg
~about the $\hat{z}$ axis.
Since the instrument remains symmetric
from side to side ($-\hat{y}$ to $+\hat{y}$),
the rotated concentrator has
two identical ``top'' walls 
and two identical``bottom'' walls,
with the top and bottom figures
described by different elliptical sections.
The detectors are mounted 
behind the square exit aperture
formed by the four walls,
and are necessarily aligned with the concentrator walls.
In terms of the instrument coordinate system,
we may write the detector polarizations as
\begin{equation}
\hat{u} = \frac{ \hat{x} - \hat{y} }{\sqrt{2}}	
~~~~~~~~
\hat{v} = \frac{ \hat{x} + \hat{y} }{\sqrt{2}}	
\label{uv_def}
\end{equation}
with one detector sensitive to $\hat{u}$ 
rotated 45\deg ~counter-clockwise from the vertical ($+\hat{x}$)
and the other sensitive to $\hat{v}$ 
rotated 45\deg ~clockwise
(Fig. \ref{uv_schematic}).
Since the detector $[uv]$ coordinates
are symmetric with respect to any
top--bottom ($\hat{x}$) asymmetry,
the polarized beam patterns
from the detector pair
are correspondingly symmetric.
\footnote{
A previous version\cite{pixie_feed_josa} 
of the concentrator
did not employ this 45\deg ~rotation
and had a poorer match between the beams.
}

%
\begin{table}[t]
\caption{Measurement frequencies and number of modes}
\label{freq_table}
\begin{center}
\begin{tabular}{|c|c|c|}
\hline
\rule[-1ex]{0pt}{3.5ex}  Measurement & Sky & Number of Modes\\
\rule[-1ex]{0pt}{3.5ex}  Frequency & Frequency & $N_{\rm mode}$  \\
\hline
\rule[-1ex]{0pt}{3.5ex}  10.8 GHz	& 32 GHz	&	5 \\
\hline 
\rule[-1ex]{0pt}{3.5ex}  35 GHz		& 105 GHz	&	49 \\
\hline 
\rule[-1ex]{0pt}{3.5ex}  91 GHz		& 273 GHz	&	330 \\
\hline 
\end{tabular}
\end{center}
\end{table} 

\section{Polarized Beam Pattern Measurements} \label{sect:measurements}   

We measured the co-polar and cross-polar beam patterns of a scaled version of the
PIXIE concentrator. Measurements took place within  the Goddard Electromagnetic
Anechoic Chamber (GEMAC) facility  and are similar to  previous measurements
described in [\citenum{pixie_feed_josa}]. Briefly, we transmit power in a single
linear polarization and use an unpolarized Thomas Keating THz absolute power
meter to record the power at the concentrator as a function of angle from the
point-source transmitter. A wire grating between the power meter and the  exit
aperture of the concentrator defines the polarization state at the power meter.
The grating can be rotated to sample either the $\hat{u}$ or $\hat{v}$
polarization. An absorbing iris at entrance aperture of the concentrator
truncates the corners of the aperture and serves as a beam stop to circularize
the beam patterns. 

Although PIXIE's FTS has frequency
response to THz frequencies, the fixed noise of the power meter and  the limited
broadcast power available at frequencies above 100 GHz make direct measurements
at higher frequencies impractical. Instead, we increase all dimensions of the
PIXIE concentrator by a factor of three, and measure the beam patterns of the
scaled concentrator at frequencies  10.8 GHz, 35 GHz, and 91 GHz corresponding
to sky measurements of 32 GHz, 105 GHz, and 273 GHz for the un-scaled
concentrator. The frequencies are selected to measure the beam pattern in the
few-mode limit, the multi-moded geometric optics limit, and an intermediate case
(Table \ref{freq_table}). Note that the frequency spectrum of the CMB polarization
follows the derivative $dB/dT$ of a blackbody spectrum, and peaks at frequencies
near 270 GHz well into multi-moded operation. GEMAC data at 91 GHz sample the
beam pattern near this peak.

\begin{figure}[b]
\centerline{
\includegraphics[width=4.0in]{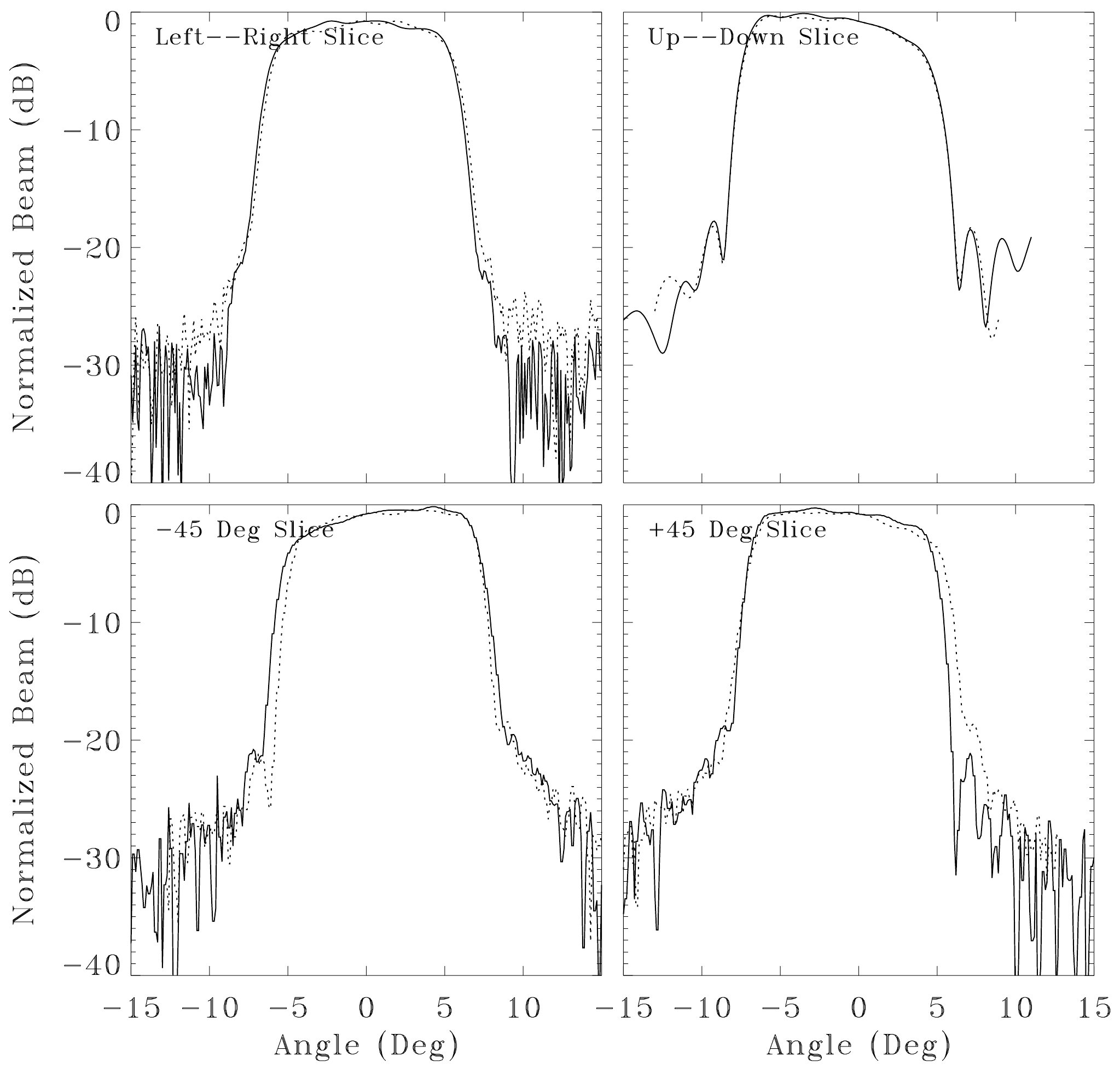}
}
\caption[Co-polar beam pattern for $\hat{u}$ and $\hat{v}$ polarization]
{
Comparison of co-polar beam patterns 
for $\hat{u}$ (solid line) and $\hat{v}$ (dotted) polarization
measured at 35 GHz for the scaled concentrator.
Four slices through the flat-topped 2-dimensional beam pattern are shown.
Left--right and up--down slices (top row)
correspond to slices oriented along the instrument 
$\hat{x}$ and $\hat{y}$ coordinates.
$+45\deg$~and $-45\deg$~slices (bottom row)
correspond to the detector
$\hat{u}$ and $\hat{v}$ coordinate directions
(see Figure \ref{pixie_schematic}).
Mounting the concentrator at a 45\deg ~angle
produces nearly identical beam patterns
for both detector polarizations.
}
\label{copol_slice_30}
\end{figure}

The measured beam patterns 
include the response of the
concentrator as well as the detector.
The PIXIE detectors 
use silicon wires 
degeneratively doped with phosphorus 
to absorb light in a single linear polarization
\cite{nagler/etal:2016}.
The measured beam patterns use
a wire grating
to reflect light in a single linear polarization,
then measuring the transmitted light
in the orthogonal polarization 
using an unpolarized power meter.
The geometry of the
power meter aperture and its location within the integrating cavity,
as well as the position of the reflective wire grating,
are appropriately scaled
to match the detector for the PIXIE FTS.

Figure \ref{copol_slice_30} shows the 
co-polar beam patterns 
measured at 35 GHz with the scaled concentrator
(corresponding to 105 GHz for the unscaled design).
The top panel shows slices through the flat-topped beam
in the left--right 
(instrument $+\hat{y}$ to $-\hat{y}$)
and up--down
(instrument $+\hat{x}$ to $-\hat{x}$)
directions.
As expected,
the beam patterns 
are symmetric with respect to $\hat{y}$
but show a few-dB ``tilt'' in $\hat{x}$
from the off-axis illumination.
The bottom panel 
shows the beam patterns
along $-45\deg$~and $+45\deg$~slices (bottom row)
correspond to the detector
$\hat{u}$ and $\hat{v}$ coordinate directions.
Note that for all four slices,
the co-polar beam pattern sampled by the $\hat{u}$ detector
is nearly identical to the pattern 
sampled by the orthogonal $\hat{v}$ detector.
This minimizes temperature--polarization mixing
between unpolarized (temperature) gradients
and the smaller polarized sky signal.

\begin{figure}[b]
\centerline{
\includegraphics[width=4.5in]{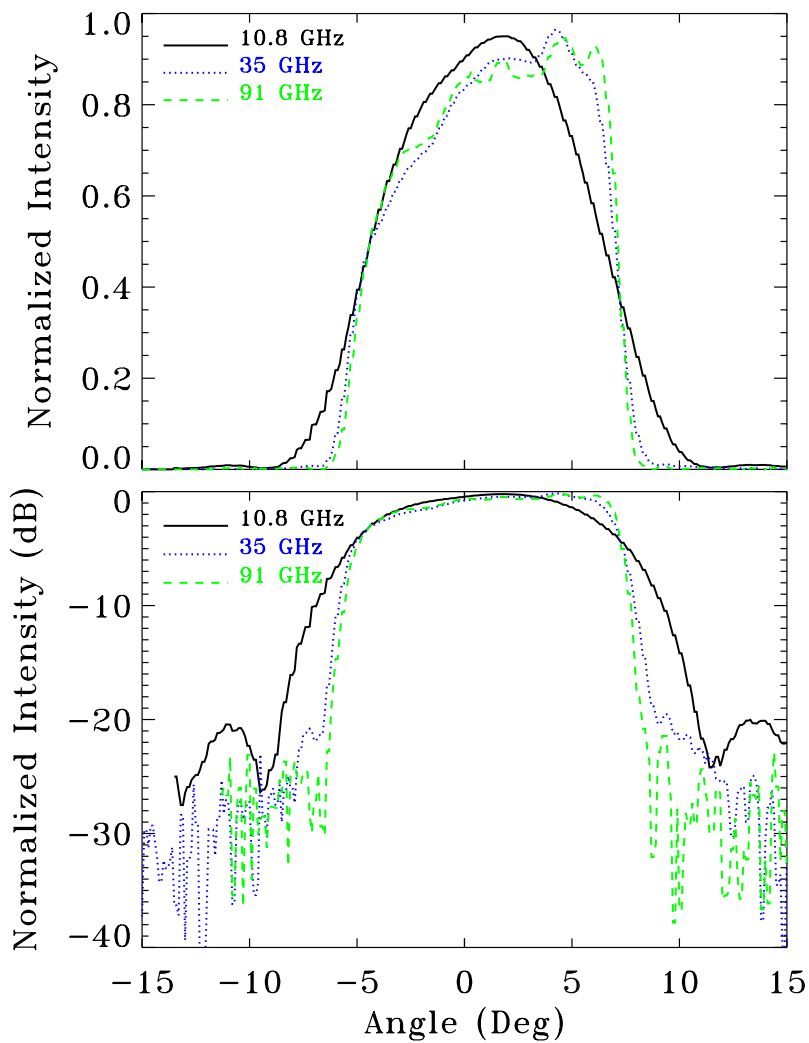}
}
\caption[Co-polar beam pattern vs frequency]
{
Comparison of co-polar beam patterns 
as a function of frequency
for the $\hat{u}$ polarization
along the $+45\deg$ slice.
Beam patterns are shown for both a linear scale (top plot)
and logarithmic scale (bottom plot).
Diffraction broadens the beam pattern at the lowest frequency,
but the beam shape is nearly constant
at higher frequencies.
}
\label{copol_3_freq}
\end{figure}

Figure \ref{copol_3_freq}
compares a slice through the copolar $\hat{u}$ beam
in the $\hat{u}$ orientation
at each of the mapped frequencies.
The beam profile is similar at 35 and 91 GHz
but broadens at 10.8 GHz
where diffraction plays a more significant a role.

\begin{figure}[t]
\centerline{
\includegraphics[width=5.0in]{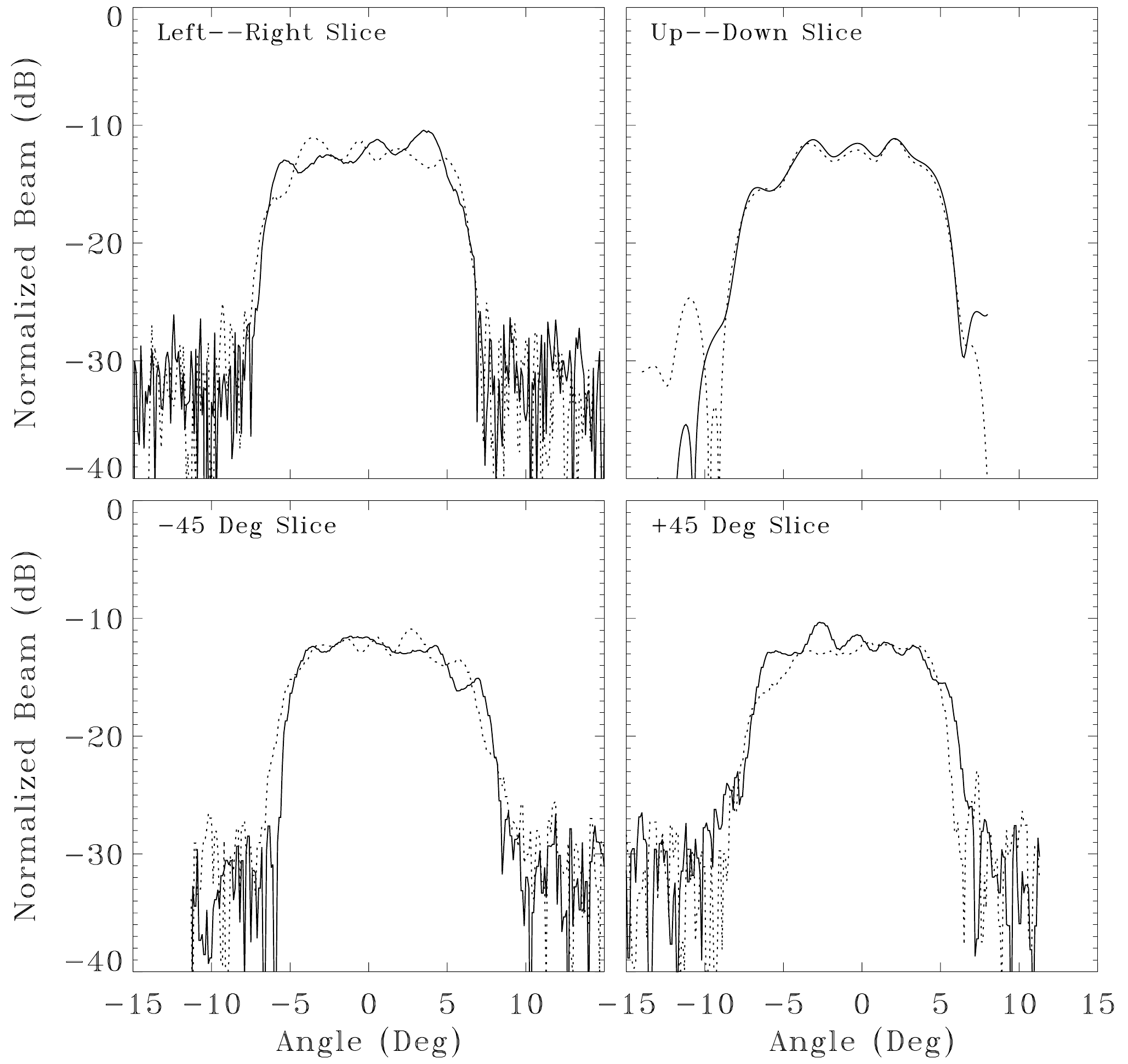}
}
\caption[Cross-polar beam pattern for $\hat{u}$ and $\hat{v}$ polarization]
{
Comparison of cross-polar beam patterns 
for $\hat{u}$ (solid line) and $\hat{v}$ (dotted) polarization
measured at 35 GHz for the scaled concentrator.
Beam slices are the same as Figure \ref{copol_slice_30}.
}
\label{xpol_slice_30}
\end{figure}

The co-polar beam patterns measure the response 
of the concentrator
to a linearly polarized plane wave,
when the polarization of the incident beam
matches the polarization accepted by the detector
(e.g. $\hat{u}$ incident polarization
measured by the $\hat{u}$ detector).
We also measure the cross-polar response
when the detector is orthogonal to the incident polarization
($\hat{u}$ incident polarization
measured by the $\hat{v}$ detector).
Figure \ref{xpol_slice_30} shows the cross-polar beam pattern
measured at 35 GHz.
As with Fig \ref{copol_slice_30},
we show the cross-polar response
in 4 slices through the flat-topped beam,
corresponding to the
principal instrument  
($\hat{x}$ and $\hat{y}$)
and detector
($\hat{u}$ and $\hat{v}$)
symmetries.

The cross-polar beam pattern has a similar flat-topped shape
as the co-polar beam,
but at a reduced amplitude.
The dominant effect of the cross-polar response
is a modest reduction in polarization efficiency
(Table \ref{pol_eff_table}).
The measured beam patterns
correspond to polarization efficiency
of $\sim$0.95, nearly independent of frequency.
The polarization efficiency reduces the instrument response 
to a linearly polarized sky signal,
and may be absorbed into the instrument calibration.
It thus affects instrument signal-to-noise estimates,
but does not induce systematic error
in the measured sky polarization.

\section{Comparison with Model}
\label{sect:model} 

We model the co-polar and cross-polar beam patterns
using a ray-trace code in the geometric optics limit.
As discussed in [\citenum{pixie_feed_josa}],
we account for operation in the few-mode limit
by first binning the modeled beam pattern on a rectangular grid,
Fourier transforming the binned beam map,
sorting the complex Fourier coefficients by angular frequency,
and Fourier transforming back to real space
using only the lowest $N_{\rm mode}$ values.
This removes the high spatial frequency information
from the modeled beam pattern,
approximating the effect of few-mode operation
without computationally expensive phase matching.
We then convolve the mode-truncated pattern
with the Airy pattern for the appropriate observing frequency
to approximate the effects of diffraction at the iris.

%
\begin{table}[t]
\caption{Polarization efficiency}
\label{pol_eff_table}
\begin{center}
\begin{tabular}{|c|c|}
\hline
\rule[-1ex]{0pt}{3.5ex}  Measurement &  Polarization\\
\rule[-1ex]{0pt}{3.5ex}  Frequency & Efficiency  \\
\hline
\rule[-1ex]{0pt}{3.5ex}  10.8 GHz	& 0.96	\\
\hline 
\rule[-1ex]{0pt}{3.5ex}  35 GHz		& 0.93	\\
\hline 
\rule[-1ex]{0pt}{3.5ex}  91 GHz		& 0.95	\\
\hline 
\end{tabular}
\end{center}
\end{table} 

\begin{figure}[b]
\centerline{
\includegraphics[width=5.0in]{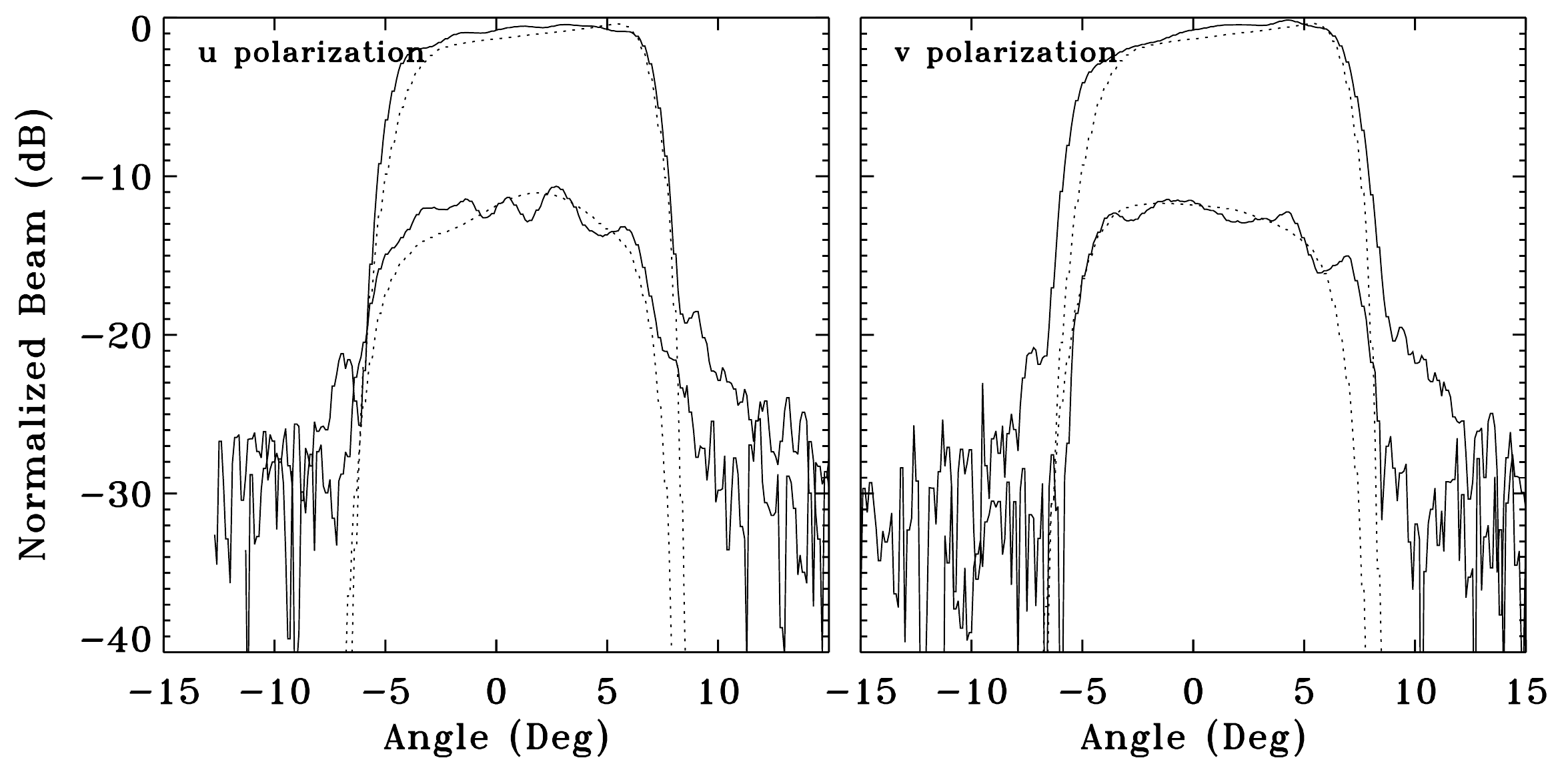}
}
\caption[Measured vs model beam patterns]
{
Comparison of measured beam patterns at 35 GHz (solid lines)
to the model including effects of diffraction 
and the number of modes (dotted lines).
The upper set of curves show the co-polar beam
while the lower set shows the cross-polar response.
The primary effect of the cross-polar response
is a 5\% loss in polarization efficiency.
}
\label{data_vs_model}
\end{figure}

Figure \ref{data_vs_model} compares the 
co-polar and cross-polar beam patterns
at measured frequency 35 GHz.
Data and model agree well across the flat-topped beam,
with rms difference 0.7 dB
in both the co-polar and cross-polar response
for either $\hat{u}$ or $\hat{v}$ polarization.
The agreement between the
measured and modeled beam patterns
highlights the utility of the simple model technique.
Direct electromagnetic computation of beam patterns
through software code such as CST or HFSS
requires amplitude and phase matching
for each mode.
A large number of modes
in an off-axis geometry
requires long computational times
for a single configuration,
becoming prohibitive
for a design process
with multiple iterations.
The ray-trace code is simple, fast,
and readily accounts for
effects
including off-axis orientation
and
complex detector geometries
while retaining 
accurate modeling
of both the co-polar and cross-polar beam response.

\section{Discussion}
\label{sect:discussion}  

Each PIXIE detector samples a single linear polarization
(Stokes $Q$ in instrument-fixed coordinates).
The instrument rotates about the beam boresight
to modulate the sky signal,
allowing reconstruction of Stokes $Q$ and $U$ parameters
for each independent pixel on the sky
\cite{kogut/etal:2011}.
Structure within the cross-polar response
on angular scales smaller than the beam width
can generate systematic error in the reconstructed sky signal.
We estimate the magnitude of this effect
by convolving the measured co-polar and cross-polar beam patterns
with Monte Carlo simulations of polarized CMB sky signals.

The beam patterns for the full PIXIE instrument
depend on both the concentrator
and the fore-optics.
Ray-trace modeling shows that the cross-polar response
is dominated by the concentrator.
We thus use the beam patterns measured for the
scaled concentrator to estimate the
systematic error resulting from the 
cross-polar beam response of the full instrument.
The concentrator is designed to illuminate
the transfer mirror T5
coupling the detectors to the spectrometer.
The measured beam widths of the concentrator thus do not represent
the width of the final PIXIE beams on the sky.
However, since the optics conserve beam etendu,
we may approximate the final beam on the sky
by scaling the measured co-polar and cross-polar beam patterns
by the diameter ratio of the 112 mm coupling mirror
used for beam measurements
to the PIXIE 550 mm primary mirror.
The resulting beam has a diameter of 2.6\deg.

We generate random realizations
of CMB polarization
using a standard $\Lambda$CDM cosmology
\cite{planck_cosmology_2016},
evaluated to harmonic moment
$\ell = 1500$
(angular scale 7\amin).
We convolve the simulated Stokes $Q$ and $U$ maps 
with the scaled beam
to generate the instrument response at fixed spin angle $\gamma$,
then rotate the beam with respect to the sky
and repeat for 32 angular steps
uniformly covering spin angle [0, $2\pi$].
Although the PIXIE fore-optics and FTS interfere the signals
from two co-pointed beams,
we ignore any additional cancellation of beam effects
from the fore-optics
and simply scale the beam patterns
measured from the concentrator.
The resulting signal is thus an upper limit
to the systematic error expected from the full PIXIE optical system.

\begin{figure}[b]
\centerline{
\includegraphics[width=4.0in]{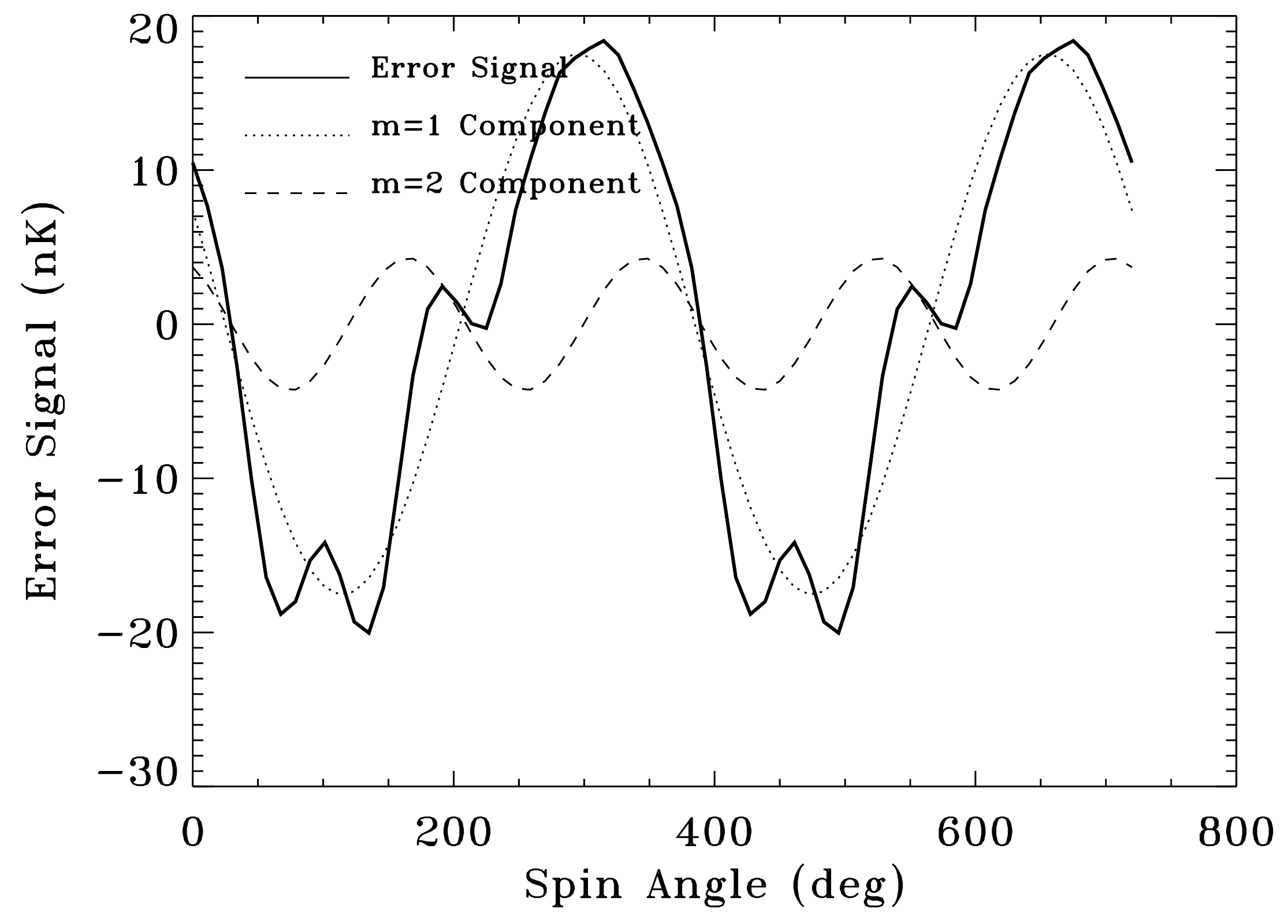}
}
\caption[Systematic error from cross-polar response]
{
Systematic error for the 35 GHz cross-polar response
convolved with simulated CMB maps.
The cross-polar response (solid line)
is dominated by an $m=1$ dipole component (dotted line)
which does not contribute to the reconstructed sky polarization.
The $m=2$ component (dashed line)
creates a systematic error in polarization,
but is small compared to the $r=0.01$ inflationary signal.
}
\label{xpol_syserr}
\end{figure}

Figure \ref{xpol_syserr} shows the resulting 
systematic error signal
for two complete rotations of the beam over the simulated sky.
The solid line shows the effect
of the cross-polar response of the scaled 35 GHz beam
(corresponding to 105 GHz sky signal).
The ``tilt'' in the beam from the off-axis illumination
modulates the response at the spin period.
Since a true polarization signal
is modulated twice per spin,
the tilt does not contribute to the reconstructed polarization.
We fit the simulated data
to azimuthal spin modes,
\begin{equation}
T(\gamma) = \sum_m \left( a_m \cos(m \gamma) + b_m \sin(m \gamma) \right)
\label{m_eq}
\end{equation}
where $m$ denotes the azimuthal order.
The dotted line shows the dipole ($m=1$) component.
The $m=2$ component (dashed line)
is degenerate with true sky polarization,
but is substantially smaller.
We repeat the analysis for 100 realizations of the CMB sky.
The $m=2$ systematic error component has mean amplitude
$\left< ~(a_2^2 + b_2^2)^{1/2} ~\right> = 3.3$ nK 
with standard deviation 1.7 nK.
Compared to the inflationary B-modes,
the $m=2$ error from the measured cross-polar response
corresponds to mean 
$r = 1.2 \times 10^{-3}$
with standard deviation
$r = 0.6 \times 10^{-3}$.
Note that this represents an upper limit to the
potential systematic error before correction.
CMB polarization is dominated by the E-mode component.
The measured cross-polar beam pattern 
may be convolved with maps of the E-mode polarization
to correct the cross-polar response,
further reducing its effect on B-mode searches.

\section{Conclusion}
\label{sect:conclusion}  

We measure the co-polar and cross-polar beam patterns
from a multi-moded concentrator 
designed for the PIXIE polarimeter.
The beams provide a flat-topped pattern on the sky
and are fully symmetrized between orthogonal linear polarizations.
The cross-polar beam pattern
has similar angular dependence as the co-polar beam;
the dominant effect of the cross-polar response
is a reduction in polarization efficiency 
to a measured efficiency of 95\%.
Structure in the cross-polar response
on angular scales smaller than the 2.6\deg~
co-polar beam
creates a systematic error when reconstructing the
polarization signal from the sky.
Monte Carlo simulations
show the un-corrected error signal
to have mean amplitude $3.3 \pm 1.7$ nK,
corresponding to an error
$\delta r = [ 1.2 \pm 0.6 ] \times 10^{-3}$
for the inflationary B-mode signal.
Convolving the measured cross-polar beam pattern
with maps of the CMB E-mode polarization
provides a template for correcting the cross-polar response,
reducing it to negligible levels.

\acknowledgments 
The authors gratefully acknowledge the assistance of
G. De Amici and S. Seufert
for measurements of the polarized beam patterns.


\vspace{2ex}
\noindent
\textbf{Alan Kogut} is an astrophysicist at NASA's Goddard Space Flight Center.
He received his A.B. from Princeton University
and his PhD from the University of California at Berkeley.
His research focuses on observations of the
frequency spectrum and linear polarization
of the cosmic microwave background
and
diffuse astrophysical foregrounds
at millimeter and sub-mm wavelengths.

\vspace{2ex}
\noindent
\textbf{Dale Fixsen} is an astrophysicist at the University of Maryland. 
He received his B.S. from Pacific Lutheran University, Tacoma Washington
and his PhD from Princeton University. 
His research focuses on the spectrum, temperature, and polarization 
of the cosmic microwave background 
and the radio and infrared cosmic backgrounds.


\end{spacing}
\end{document}